\documentclass{llncs}
\usepackage{epsfig}
\usepackage{color}
\usepackage{subfigure}
\usepackage{url}
\usepackage{tabularx}

\begin{document}

\date{}

\title{\Large \bf Toward automatic censorship detection in microblogs}

\author{
{\rm Donn Morrison}\\
\institute{
Department of Computer and Information Science\\
Norwegian University of Science and Technology\\
Trondheim, Norway\\
donn.morrison@idi.ntnu.no}
} 

\maketitle

\thispagestyle{empty}

\subsection*{Abstract}

Social media is an area where users often experience censorship through a variety of means such as the restriction of search terms or active and retroactive deletion of messages. In this paper we examine the feasibility of automatically detecting censorship of microblogs. We use a network growing model to simulate discussion over a microblog follow network and compare two censorship strategies to simulate varying levels of message deletion. Using topological features extracted from the resulting graphs, a classifier is trained to detect whether or not a given communication graph has been censored. The results show that censorship detection is feasible under empirically measured levels of message deletion. The proposed framework can enable automated censorship measurement and tracking, which, when combined with aggregated citizen reports of censorship, can allow users to make informed decisions about online communication habits.

\section{Introduction}
\label{sec_intro}

The recent and continuing popularity of the social aspect of the Internet, in particular social media and online social networks (OSNs), has facilitated unprecedented new ways of communication and levels of information sharing. This new freedom has challenged many governments, organisations and businesses which, for legitimate reasons or not, are struggling to control dissemination of news events, digital content and other sensitive information. When the censorship is acknowledged, justification ranges from maintaining public order and safety \cite{dick2012established} to protection of morality from obscenity \cite{sanghun_nytimes_2012} to the protection of intellectual property or copyright \cite{stone_nytimes_2008}. In most cases, however, censorship is unquestionably a hindrance to a free and transparent society where citizens are able to participate by expressing ideas and opinions openly and without fear of reprisal.


Exposing censorship and the methods used to achieve it puts pressure on repressive elements and allows citizens to make informed decisions about how they participate in society. Projects such as Herdict \cite{physorg_herdict_2009,Hwang200715} and ConceptDoppler \cite{DBLP:conf/ccs/CrandallZBBE07} have undertaken to measure and track censorship online. However, Herdict relies on user reports and neither Herdict nor ConceptDoppler focus on user communication in OSNs.


Recent research has focused on identifying sensitive keywords as well as influential or controversial users who are more likely to be censored \cite{fm2012,DBLP:journals/corr/abs-1211-6166,DBLP:journals/corr/abs-1303-0597}. However, these entities cannot always be anticipated before the censorship occurs, and therefore it is desirable to focus on features that are not content-based.

This paper proposes a novel method for censorship detection that does not rely on keyword lists or other forms of content. Instead, the approach classifies communication graphs derived from OSNs based solely on topological properties. More specifically, this work makes the assumption that user communication behaviour on microblogs, i.e., the posting and replying of messages, is generated by a random process that can be approximated using a graph generator. More importantly, however, is the further assumption that acts of censorship, specifically message deletion, results in a definite and measurable effect on the communication graphs.

Combined with the aforementioned citizen reports of censorship, OSNs and other online communities that deviate from known norms could be flagged as being censored and users could be warned to adapt strategies for organising and disseminating information.

This paper contributes to the understanding of the effects of social media censorship on network structure in the following two ways:

\begin{enumerate}
\item We identify salient topological features and show how they are affected by varying levels of censorship;
\item We propose a framework for automatic censorship detection at the network level that is content-agnostic.
\end{enumerate}

In light of recent political events such as the Arab Spring of 2011, the current conflict in Syria and the ongoing censorship of media in China, there is an urgent need for a framework for the measurement of censorship online to ensure freedom of speech and access to information, and, in the larger context, maintain a free and open Internet. The approach aims to fill this gap and in doing so facilitate a better understanding of network censorship and ultimately provide a means for automated measurement, tracking and monitoring of censorship on the Internet.

The remainder of the paper is organised as follows. First we highlight related work that examines censorship of microblogs, primarily on the Sina Weibo network popular in China. Then in Section \ref{sec_metho} we present the methodology which outlines the data generation, graph feature extraction and classifier setup. Finally, Section \ref{sec_results} presents the results and discussion.

\section{Related work}
\label{sec_related}

Censorship in the context of social media has been defined as the ``suppression, limiting or deleting of objectionable'' content or any other form of speech or expression \cite{deibert2008-black,dick2012established}. There have been numerous works documenting instances and trends of censorship and circumvention strategies online, generally anecdotal or qualitative in nature, often relying on first-hand accounts \cite{roberts2011circumvention,citeulike:10435050,sanghun_nytimes_2012}. However, some recent research has employed quantitative methods to measure and compare censorship practices on different OSNs \cite{fm2012,DBLP:journals/corr/abs-1211-6166,DBLP:journals/corr/abs-1303-0597}.

Detection of deleted posts has been used in previous work to quantify censorship. The most pervasive methodology involves sampling microblog posts over a period of time to capture sensitive political events while querying the service at regular intervals to determine if any of the posts have been deleted \cite{fm2012,DBLP:journals/corr/abs-1211-6166,DBLP:journals/corr/abs-1303-0597}.

Bamman et al. \cite{fm2012} uncovered politically sensitive terms more likely to be actively and retroactively deleted in a comparison between censorship on Twitter and China's Sina Weibo microblogging services. A random sample of collected messages found that 16.25\% were deleted from the Weibo network. Geographic distribution was found to have a strong impact on message deletion rates, with up to 53\% of sampled messages originating from some Chinese provinces deleted.

Initial research by \cite{DBLP:journals/corr/abs-1211-6166} shows that active and retroactive censorship to a large extent succeeds in stemming the spread of information on microblogs. In a subsequent work, the authors studied the time distribution of deleted messages and found that nearly 30\% of deletions happen in the first 5-30 minutes and up to 90\% of deletions occur within 24 hours of the posting \cite{DBLP:journals/corr/abs-1303-0597}. Extrapolating the sampled data to message posting rates, the authors estimated that up to 4,200 workers working eight hour shifts would be required to match the demand for censorship levels on Sina Weibo alone. Furthermore, the authors uncovered censorship behaviour such as peak hours where censorship occurs and the practice of deleting entire repost cascades started from a single sensitive post. Ultimately, a complex array of censorship practices filter the continuous stream of Weibo posts such that sensitive topics do not enter into mainstream discussion.

Network perturbation and resilience is a closely related field where network metrics are studied under destructive processes that iteratively remove nodes or edges \cite{cohen2001breakdown,gallos2004tolerance,yadav2012nexcade}, however, these works do not consider censoring models for these processes nor do they formulate the problem as one of classification.

Despite these important works, no research to date has explored the effects of censorship on the underlying structure of the network and furthermore no research exists that attempts to automatically detect and classify censorship in these networks. Given that online social networks have certain universal properties \cite{Chakrabarti2006a}, it is likely that common strategies of censorship such as limiting or deleting content or users from the network would have measurable effects on these properties. This research constitutes a first step to fill this gap by studying these effects.

\section{Methodology}
\label{sec_metho}

In this section we detail the methodology. First, we define a reply-graph over a microblog follow network. Then, we show how we use the \textit{configuration model} to generate reply-graphs and present two methods to simulate censorship of these networks. Next, we introduce topological features extracted from the reply-graph that are then used to train a support vector machine in order to classify network censorship. Finally, our experimental setup is presented.

\subsection{Definitions}

Consider a directed multigraph $G=(V,E)$ without self loops where the nodes $V$ represent users and the edges $E$ represent microblog posts over the follow network. That is, an edge $e_{ij} \in E$ if user $v_i$ is followed by user $v_j$ \textit{and} a post from $v_i$ is shown in $v_j$'s timeline. Note, an edge $e_{ij}$ does not imply an edge $e_{ji}$. This notation corresponds to the flow of information over the edges $E$. As an example, the user $v_i$, who is followed by the set of users $S$, posts a new microblog entry $m$. Then, for each $v_j \in S$, a new edge $e_{ij}$ is created in $G$, meaning that the entry $m$ was visible in the timelines of users $S$.

To remain general, we refer to the graph $G$ as the reply-graph as in \cite{Morrison2012a}, but with the constraint that an edge is only possible if there is a follow relationship between two users.

\subsection{Configuration model}

Due to the limited availability of censored microblog reply-graphs, we have chosen to generate random graphs with similar characteristics. Simulation of network data is commonly used when access to data is limited or when characteristics of the network must be carefully controlled. Since our aim is the study of reply-graphs, we make use of the directed multigraph \textit{configuration model} (CM) proposed by \cite{newman2001random} that permits random graph construction with arbitrary in and out degree distributions.

Power laws have been observed in the degree distributions of online social networks \cite{Chakrabarti2006a} although the ubiquity of data conforming to this distribution is often overstated \cite{DBLP:journals/siamrev/ClausetSN09} and depending on the network in question a closer fit may be found in any number of exponential distributions (e.g., Pareto-lognormal distribution \cite{fang2012double}). However, to simplify network generation in this preliminary work we assume the degree distributions follow a power law and generate the reply-graphs accordingly. We fix the power law exponent to $\alpha=2.0$ for both the in and out degree distributions that are used as input to the CM and set the network size to $|V|=1000$ nodes.


\subsection{Simulating censorship}

We focus on the censorship of microblog posts which are represented by the edges $E$ of $G$. That is, we do not consider the case where user accounts (the nodes $V$ of $G$) are suspended or deleted. Two censorship strategies are compared. The first is based on a uniform sampling of a fraction of the edges in $G$. This strategy can be likened to a population of users that are subjected to uniform censorship, that is, each user's post has the same probability of being deleted. This carries with it the assumption that each user is equally likely to post about a topic considered worthy of censorship, which is unlikely to be the case in the real world \cite{DBLP:journals/corr/abs-1303-0597}. Nonetheless, a uniform sampling of deleted edges serves as a useful baseline. The second strategy is based on the removal of \textit{repost} (or \textit{retweet}) cascades. Such a cascade occurs when users, upon seeing a message posted by one of their followees, choose to repost the message to their followers. The information cascades over the network according to the popularity of the original post. Censoring of entire cascades has been empirically documented in previous work \cite{DBLP:journals/corr/abs-1303-0597} where censors were shown to retroactively remove a post and all subsequent reposts. We simulate repost cascades using the independent cascade model (ICM).

\begin{figure}
\centering
\includegraphics[trim=5cm 2.5cm 2cm 1cm, clip=true, width=0.4\textwidth]{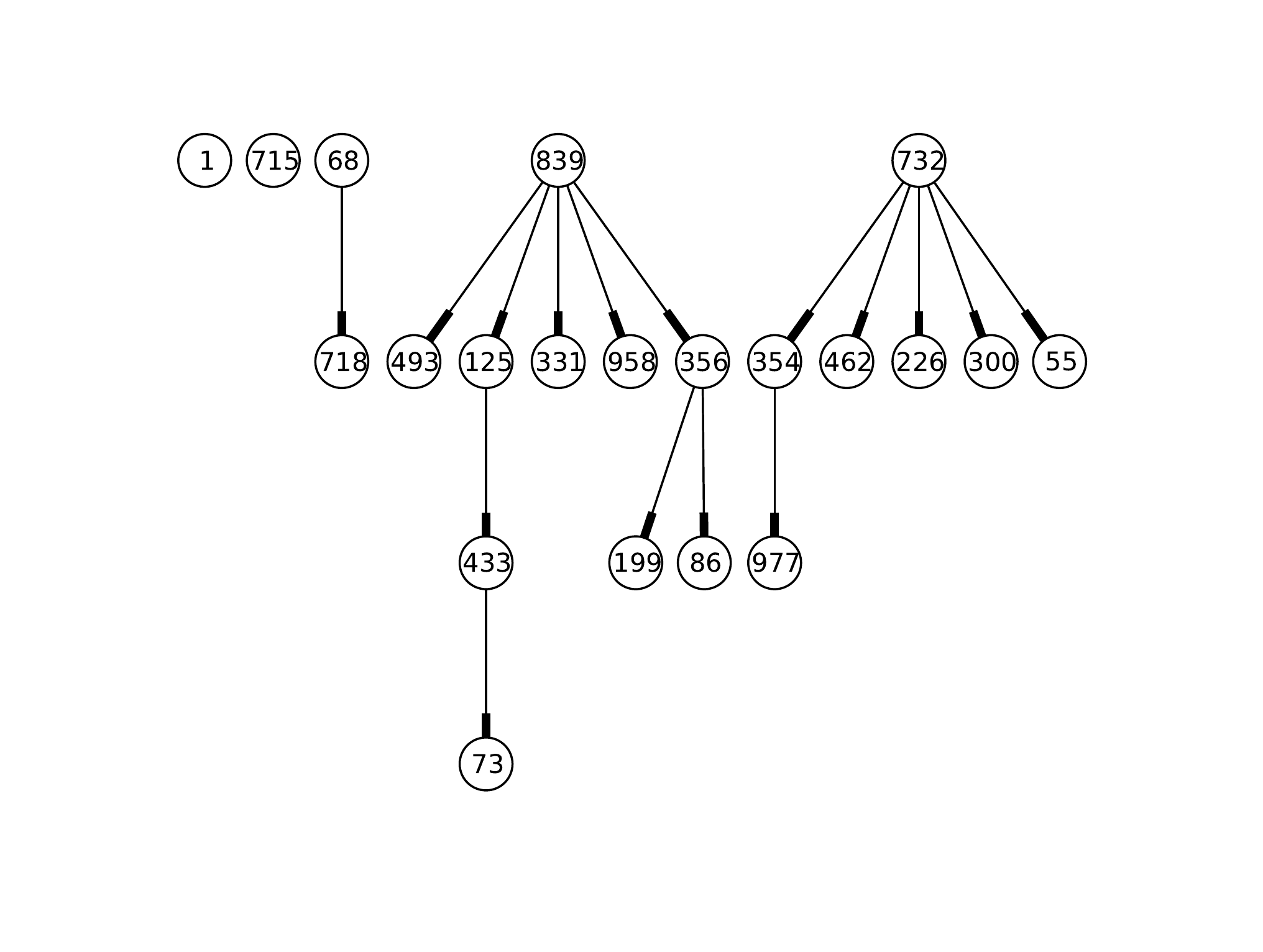}
\caption{An example set of generated repost cascades starting with three initially activated seed nodes (top row).}
\label{fig_cascade}
\end{figure}

ICM, originally proposed by Kempe \textit{et al.} \cite{Kempe2003}, is used to model spreading processes such as disease or information cascading through OSNs. The method starts with a set of activated (infected) seed nodes and at each iteration a coin is flipped to determine whether the information flows across an outgoing edge of a newly activated node. On success, the information spreads and the activated node attempts to spread the information in the next iteration. A global transmission probability is used for the coin flip and in our experiments this probability is set to 0.1 to allow for larger cascades that will form the censored edges of $G$. For simplicity, the seeds are selected by ranking the nodes according to out degree and taking the top five, corresponding to the fraction $0.005$ of the nodes $V$ in $G$. Cascades are generated with ICM until the total number of edges reaches the censorship threshold. Figure \ref{fig_cascade} shows an example set of cascades generated over $G$ using ICM.

For both the uniform and ICM-based censorship strategies, we remove a fraction $\gamma$ of the edges in $G$ for each $\gamma \in \{0.1, 0.2, 0.3, 0.4, 0.5, 0.6\}$ where the total number of edges removed equals $\gamma \times |E|$. This range allows us to study the detection of levels of censorship that have been empirically measured \cite{fm2012}.

\subsection{Network features}

Communication graphs derived from OSNs have specific characteristics distinguishing them from other networks such as random graphs \cite{Chakrabarti2006a,DBLP:journals/tkdd/LeskovecKF07,fang2012double,DBLP:journals/corr/abs-1208-2976}. For example, node degree distributions of OSNs have been shown to be exponentially distributed, following power laws \cite{Chakrabarti2006a} or Pareto-lognormal mixtures \cite{fang2012double}. OSNs also exhibit small diameters \cite{Chakrabarti2006a} that shrink with network growth \cite{DBLP:journals/tkdd/LeskovecKF07} and have a higher number of triangles\footnote{If actors A and B are connected and B and C are connected, there is a high probability that actors A and C are also connected.} compared to random graphs \cite{Chakrabarti2006a}. Centrality measures also show characteristic behaviour. In \cite{costenbader2003stability} the authors measure the stability of various centrality measures of OSNs sampled at different thresholds and show that certain centrality measures are less robust to uniform sampling.  Other research has examined spectral eigenvalue distributions for classification of biological networks as the eigenvalues are known to summarise various topological properties of graphs \cite{DBLP:journals/corr/abs-1208-2976}.

We motivate our choice of graph features by drawing on the aforementioned work. Based on these characteristics, we derive the following features and posit that they can be used to discriminate between censored and uncensored reply-graphs based on the assumption that acts of censorship fundamentally change network structure. We abbreviate feature names in parentheses for use in the figures and tables that follow.

\textbf{Average degree} (avgdeg) is defined as $\frac{2\times|E|}{|V|}$, ignoring edge direction. The \textbf{assortativity coefficient} (assort) is the degree correlation between pairs of connected nodes for the undirected equivalent of $G$. The \textbf{diameter} (dia) of $G$ is defined as the maximum shortest path for the undirected equivalent of $G$ and the \textbf{radius} (rad) is the minimum of the set of maximum path lengths from every node to every other node in the undirected equivalent of $G$.\footnote{Diameter and radius are calculated on the largest connected component.} The \textbf{average clustering coefficent} (clustering) is defined as $\frac{1}{|V|}\sum_{i\in V}{C_i}$ where $C_i$ is the local clustering coefficient measuring the number of edges divided by the number of total possible edges between the neighbours of node $i$ for the undirected equivalent of $G$. The \textbf{average betweenness centrality} (betcent) is the average of the number of shortest paths that pass through any node in $G$.

For simplicity, we assume the in and out degree distributions of $G$ to follow a power law. As such, we include the estimates of the \textbf{power law exponent} $\alpha$ (in\_alpha\_fit and out\_alpha\_fit, respectively) as well as the goodness of fit measured by the \textbf{negative log-likelihood} (in\_likelihood\_fit and out\_likelihood\_fit). The parameters are estimated by maximum likelihood estimation (MLE) as described in \cite{DBLP:journals/siamrev/ClausetSN09}. Finally, we calculate and retain the first 50 \textbf{eigenvalues of the Laplacian matrix} (spec0-49).

The resulting feature vector $F$ is of length 60 (10 topological features plus 50 Laplacian eigenvalues).

\subsection{Classification}

The classifier used in this work is the support vector machine (SVM) \cite{cortes1995support} with the radial basis function (RBF) as a kernel with parameters complexity $C=1.0$ and gamma $g=0.01$. The choice of classifier and kernel is motivated by satisfactory experimental results and the pervasive use of SVMs in machine learning literature although we note that any number of classification methods could be readily used. For brevity we omit details of SVMs and statistical learning theory and refer the reader to \cite{cortes1995support}.

\subsection{Experimental setup}

The experimental setup is as follows:

\begin{enumerate}
\item Generate $N=100$ directed multigraphs $G$ with $|V|=1000$ nodes using the CM
\item Simulate censorship uniformly and with ICM by removing $\gamma\times |E|$ edges, yielding $G_{cu}^\gamma$ and $G_{cICM}^\gamma$
\item Compute topological features $F$, $F_{cu}^\gamma$ and $F_{cICM}^\gamma$ of $G$, $G_{cu}^\gamma$ and $G_{cICM}^\gamma$, respectively
\item Classification by pairwise 10-fold cross validation on ($F$, $F_{cu}^\gamma$), ($F$, $F_{cICM}^\gamma$) with class labels $\{0, \gamma\}$
\end{enumerate}

To account for variance, Step 4 is repeated 10 times, each using a different random seed. The Java-based WEKA machine learning toolkit is used for classification and feature selection\footnote{\url{http://www.cs.waikato.ac.nz/ml/weka/}} and all experiments are conducted on an Intel quad-core i5-2520M CPU laptop running at 2.50GHz.





\section{Results and discussion}
\label{sec_results}

In this section we (1) show the effects of censorship on graph features, (2) present the classification results and (3) highlight salient graph features discovered through feature selection. Some figures have been fit with a statistical smoother and include shaded 95\% confidence intervals for readability.

\subsection{Censorship effects on graph features}

\begin{figure*}[th]
\subfigure[Assortativity]{\includegraphics[width=.33\textwidth]{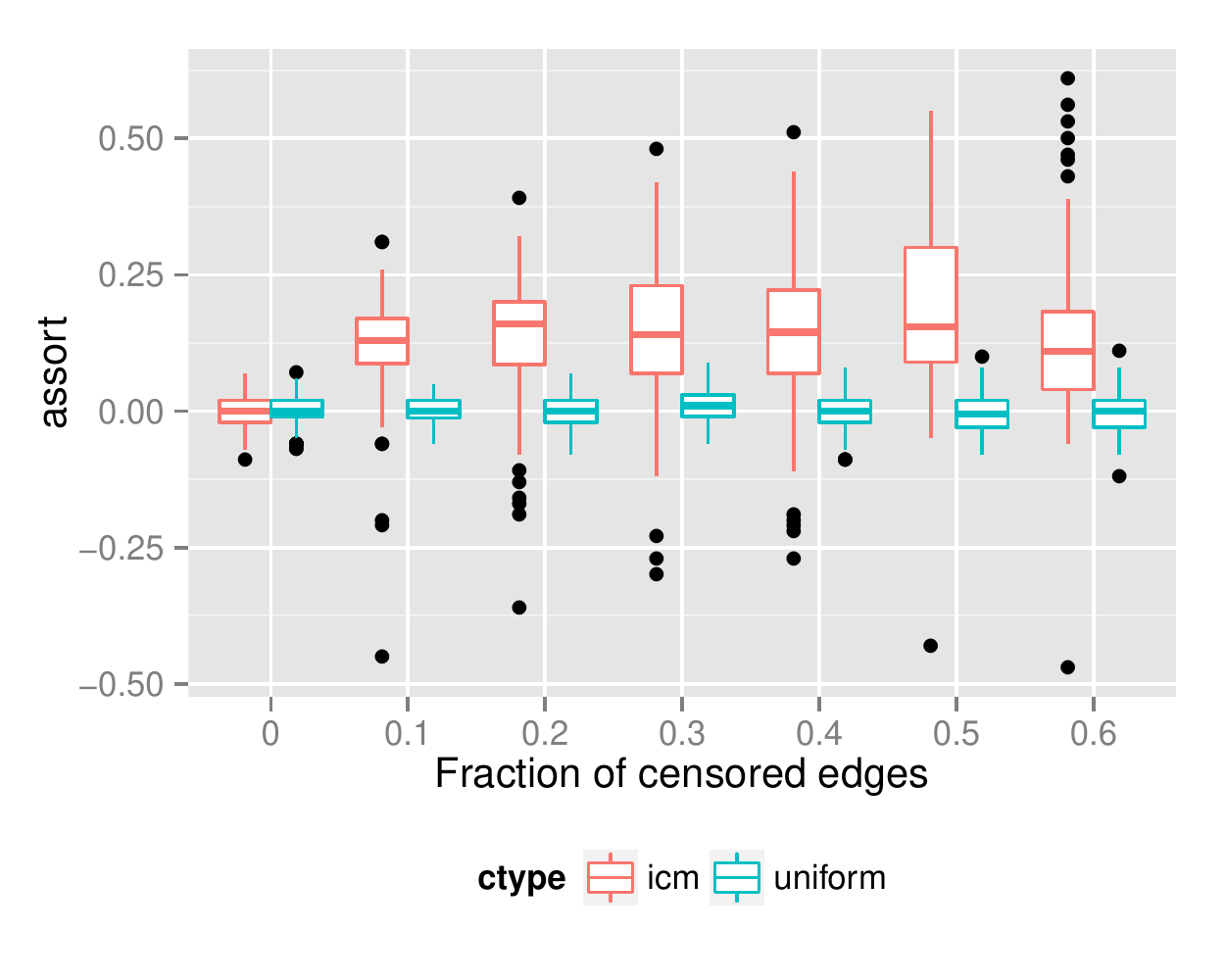}}%
\subfigure[Average clustering coefficient]{\includegraphics[width=.33\textwidth]{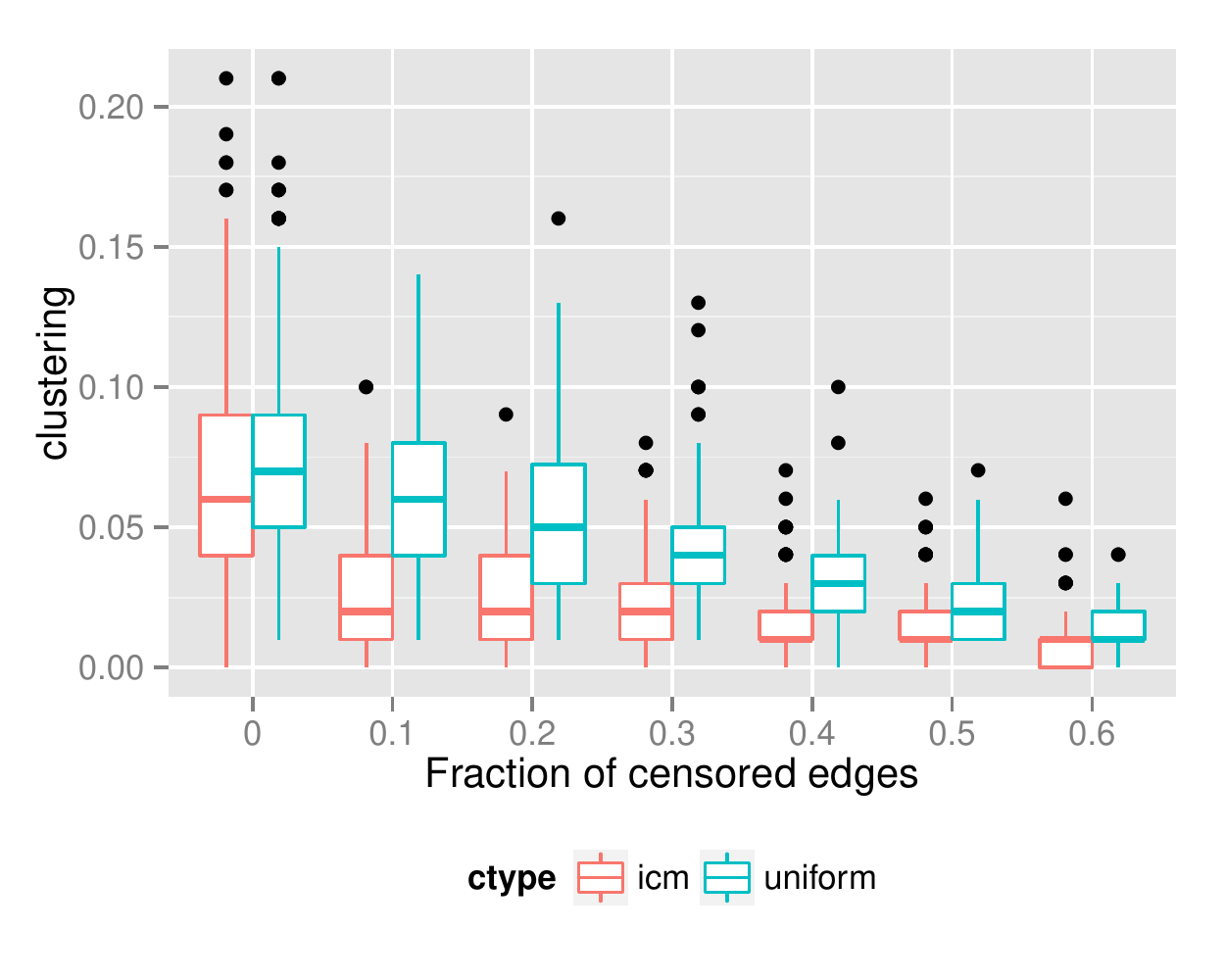}}%
\subfigure[Average degree]{\includegraphics[width=.33\textwidth]{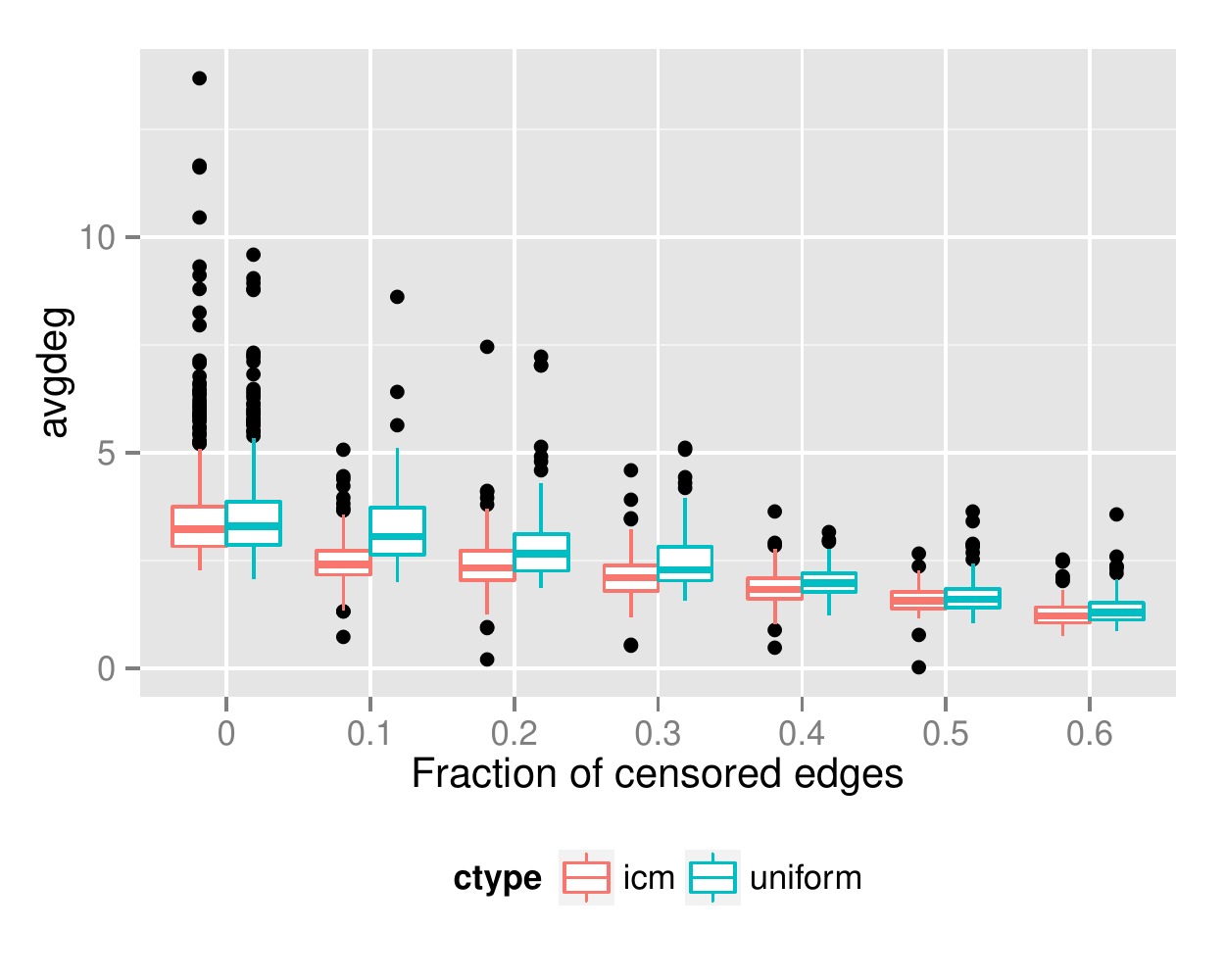}}
\subfigure[Diameter]{\includegraphics[width=.33\textwidth]{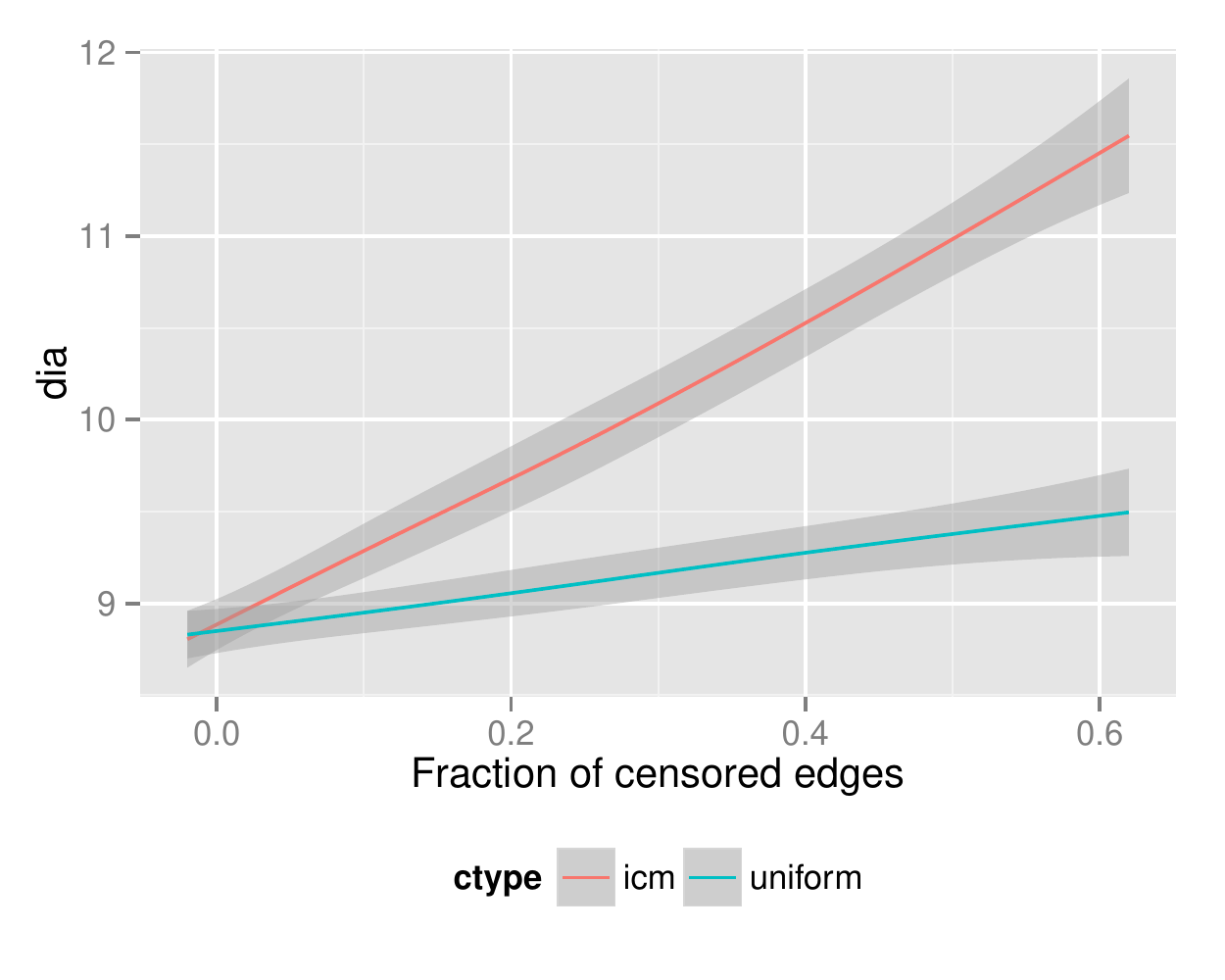}}%
\subfigure[Radius]{\includegraphics[width=.33\textwidth]{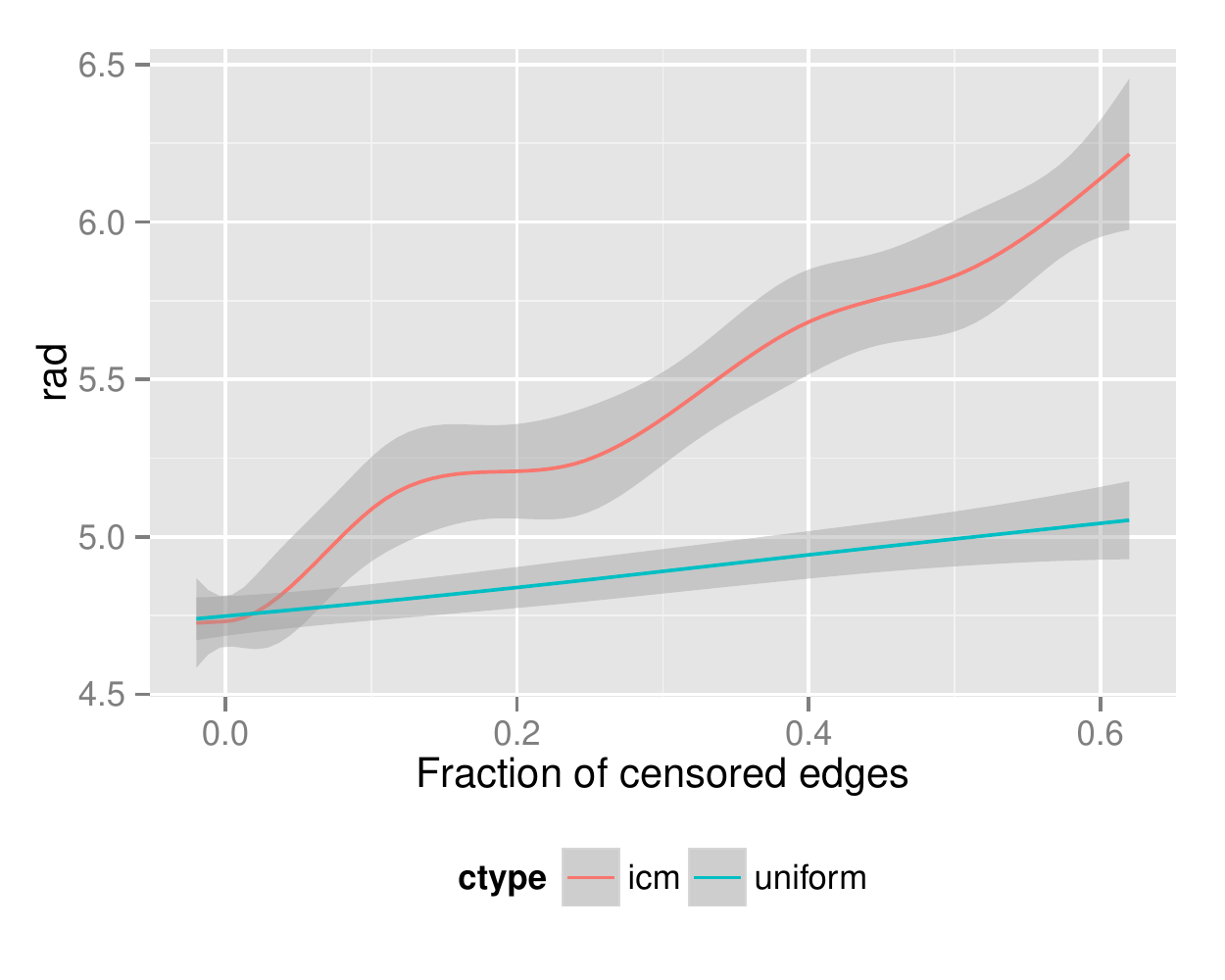}}%
\subfigure[Average betweenness centrality]{\includegraphics[width=.33\textwidth]{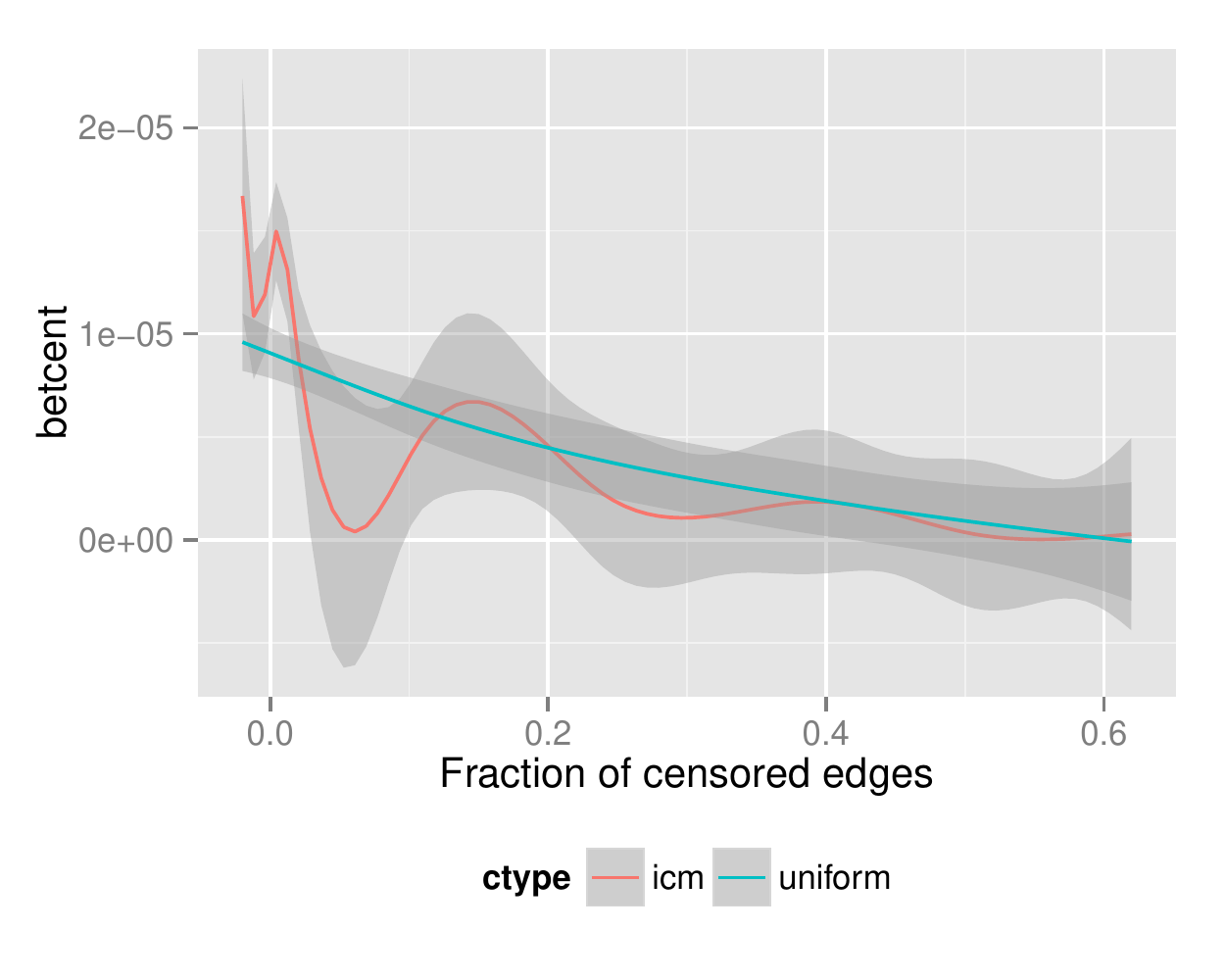}}
\caption{Topological features as a function of the fraction of censored edges.}
\label{fig_features1}
\end{figure*}

Figure \ref{fig_features1} shows the effect of censorship on the topological network features. In (a), we see that under uniform censorship, assortativity is not a discriminating feature. However, for censorship of repost cascades with ICM, we see a substantial increase in assortative nodes for mid to moderate levels of edge removal. At $\gamma=0.6$, there are less assortative nodes since at this high level much of the network structure has been lost. The average clustering coefficient (b) shows different trends for uniform and ICM. For uniform it peaks at $\gamma=0.1$ and then slowly declines while for ICM the decline is immediate with a fluctuating mean for larger values of $\gamma$. The trends for average degree (c) are similar to both strategies with a predictable decline as $\gamma$ increases. Unsurprisingly, network diameter (d) and radius (e) behave similarly because they both describe characteristics of the shortest paths. For ICM, the metrics increase with $\gamma$, consistent with previous work \cite{DBLP:journals/tkdd/LeskovecKF07}. For uniform edge removal, the growth is slower. Average betweenness centrality (f) shows a declining behaviour as $\gamma$ increases.

\begin{figure}
\subfigure[Estimated in degree $\alpha$]{\includegraphics[width=.45\textwidth]{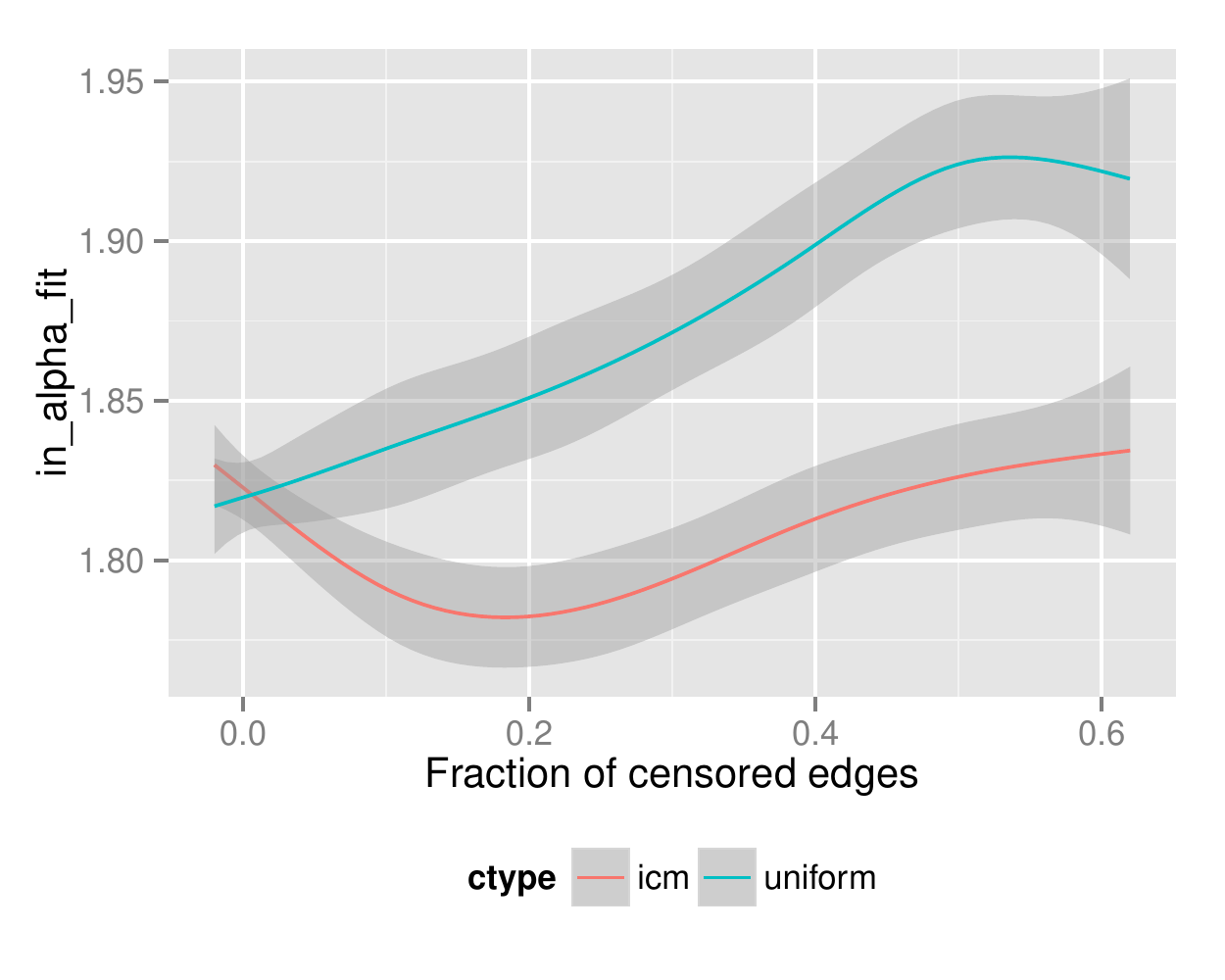}}%
\subfigure[Estimated out degree $\alpha$]{\includegraphics[width=.45\textwidth]{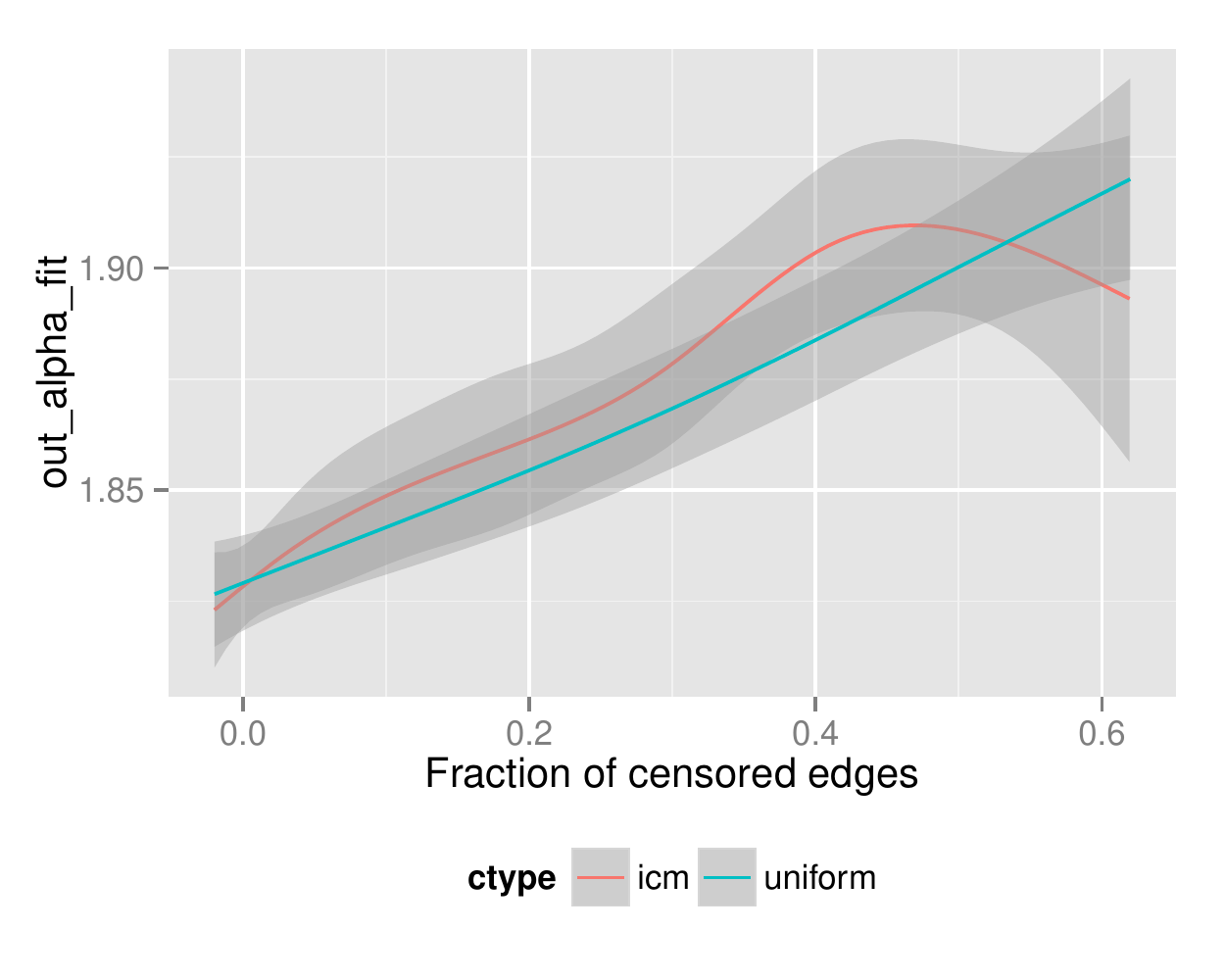}}
\subfigure[In degree $\alpha$ likelihood]{\includegraphics[width=.45\textwidth]{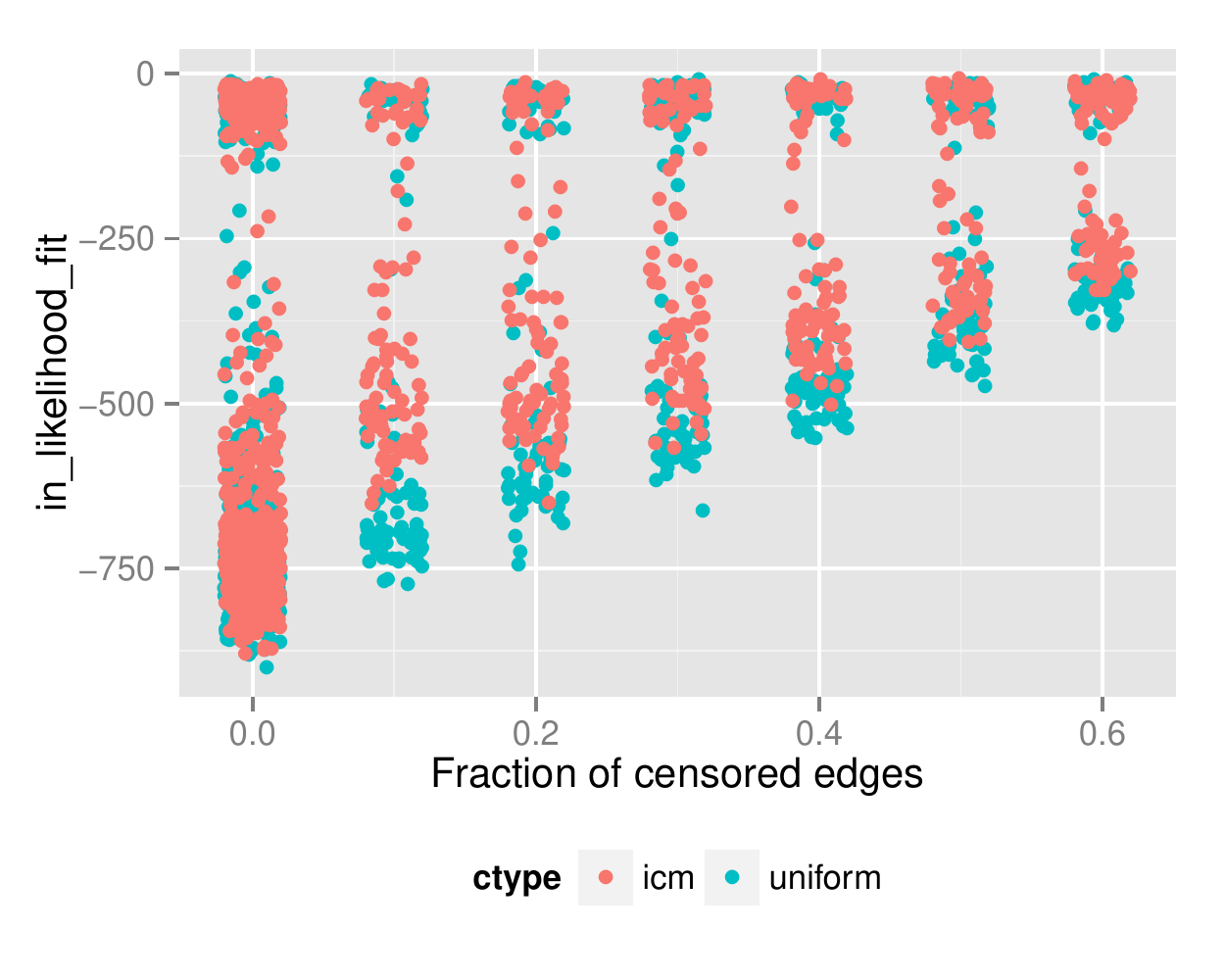}}%
\subfigure[Out degree $\alpha$ likelihood]{\includegraphics[width=.45\textwidth]{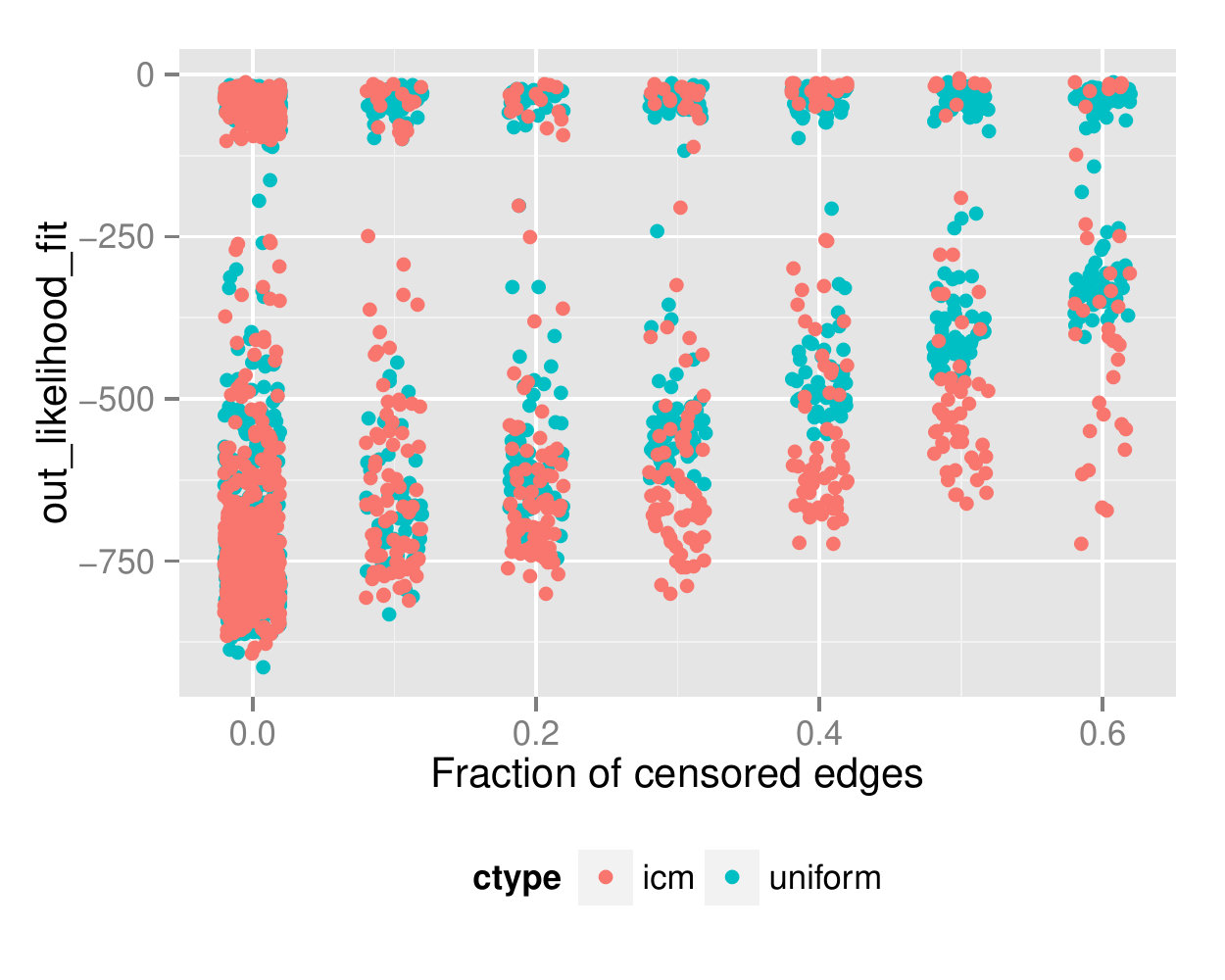}}
\caption{Estimated power law exponent $\alpha$ and log-likelihood values for in and out degree distributions as a function of the fraction of censored edges.}
\label{fig_features2}
\end{figure}


Moving our focus to the in and out degree distributions, Figure \ref{fig_features2} (a) and (b) show the estimated exponent $\alpha$ of the power law for the in and out degree distributions, respectively. Recall that the networks were generated with a value of $\alpha=2.0$. The estimated value of $\alpha$ is lower than expected, possibly due to the scale of the network and the fact that the configuration model may not accurately portray the given distributions in the resulting network. However, the estimations vary with $\gamma$, which indicates discrimination potential. The likelihoods of the power law fitting, shown in (c) and (d), are also informative features. As we deviate from the power law in the uncensored network by simulating censorship, the distributional fit becomes less and less accurate. While the power law assumption is simplistic, we expect that this can be readily generalised to other distributions if the degree sequences for uncensored networks are known a priori.




For satisfactory pairwise classification of censored versus uncensored networks, discriminating features should exhibit different values for $\gamma=0.0$ and $\gamma>=0.1$. Through visual inspection of Figures \ref{fig_features1} and \ref{fig_features2},\footnote{We omit plots for the Laplacian eigenvalues due to space considerations.} we can see that when used together, the chosen features appear to support our hypothesis that censorship, even at lower levels ($\gamma=0.1$), fundamentally alters network structure.


\subsection{Classifying censorship}

\begin{figure}[h]
\centering
\includegraphics[width=.7\textwidth]{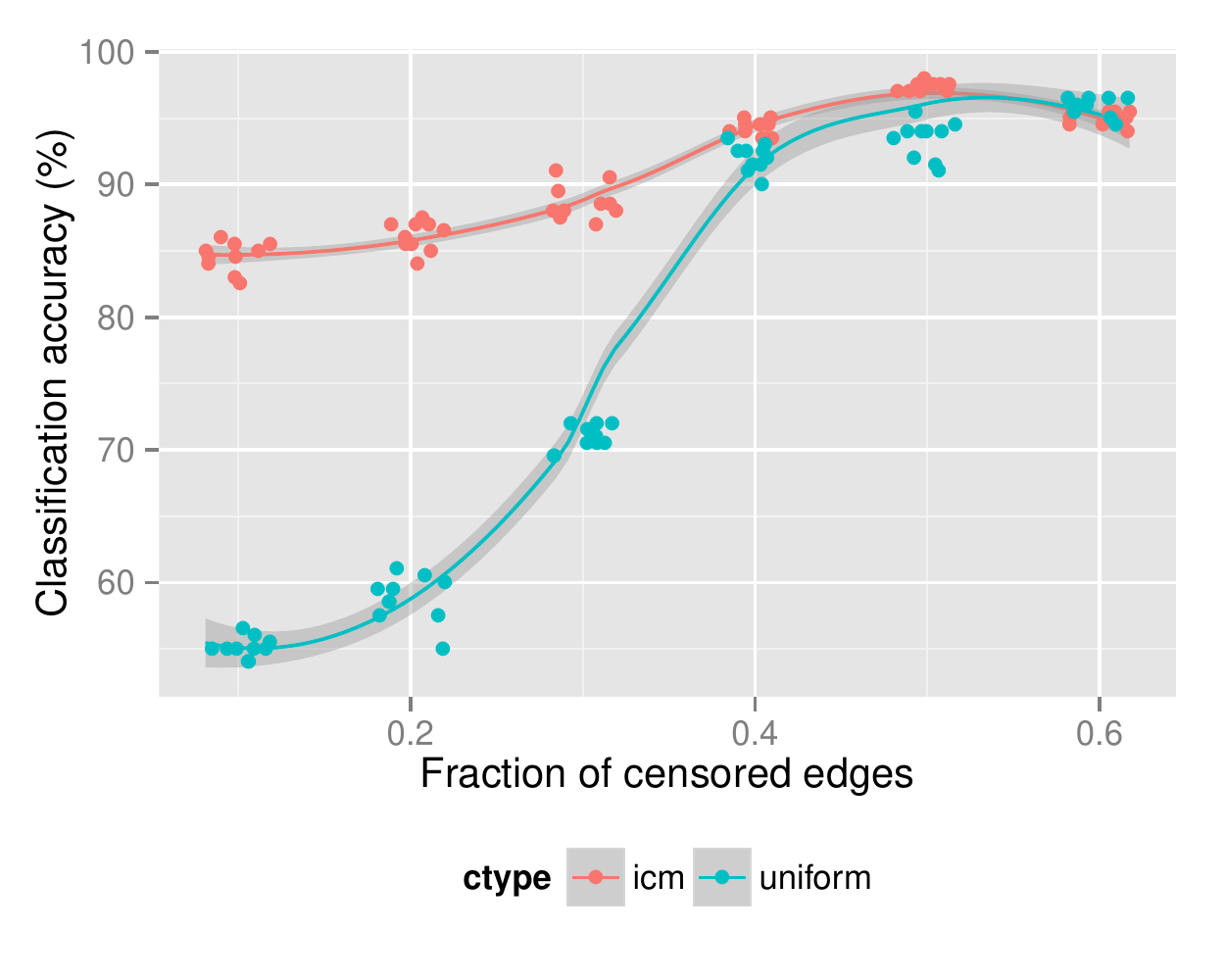}
\caption{Classification accuracy as a function of the fraction of censored edges.}
\label{fig_accuracy}
\end{figure}

Figure \ref{fig_accuracy} shows the classification accuracy of the SVM as a function of varying the fraction of censored edges for the uniform and ICM-based censorship strategies. Accuracy is substantially higher than random (50\%), and quite satisfactory for a two-class problem. We see that censorship of repost cascades (ICM) has a much stronger affect on network structure for lower values of $\gamma$ than the uniform strategy due to the inherent structural correlation between repost cascades and the reply-graph. Interestingly the accuracy plateaus to ~97\% at $\gamma=0.6$ which indicates that censorship at $\gamma>=0.5$ is trivial to detect. For uniform edge removal, there is a steep transition between $\gamma=0.2$ and $\gamma=0.4$ after which it matches classification accuracy of ICM.

\begin{table*}[ht]
\centering
\footnotesize
\caption{Feature selection results for the two censorship strategies.}
\label{tab_feature_selection}
\begin{tabularx}{\textwidth}{cr|l}
Censorship \\ strategy& $\gamma$ & Selected attributes \\
\hline
\hline
ICM & 0.1 & assort, spec2, spec3, spec38 \\
 & 0.2 & avgdeg, in\_alpha\_fit, out\_alpha\_fit, assort, spec21 \\
 & 0.3 & avgdeg, in\_alpha\_fit, assort, rad, betcent, spec1, spec13, spec29, spec35, spec48 \\
 & 0.4 & out\_alpha\_fit, assort, rad, spec1, spec12 \\
 & 0.5 & out\_alpha\_fit, assort, rad, spec0 \\
 & 0.6 & out\_alpha\_fit, assort, rad, spec2 \\
 \hline
Uniform & 0.1 &in\_alpha\_fit, spec1, spec10, spec13  \\
 & 0.2 &clustering, betcent, spec22, spec29, spec36 \\
 & 0.3 & dia, rad, clustering, betcent, spec12, spec13, spec15 \\
 & 0.4 & avgdeg, dia, rad, spec1, spec7, spec17, spec19, spec30, spec34, spec35 \\
 & 0.5 & avgdeg, dia, rad, clustering, betcent, spec0, spec5, spec9, spec18, spec28, spec32, spec45, spec47 \\
 & 0.6 & clustering, spec8
\end{tabularx}
\end{table*}

\subsection{Feature selection}

Feature selection was performed for each $\gamma$ using a greedy forward search on the entire dataset with the RBF SVM for both uniform and ICM. These results are presented in Table \ref{tab_feature_selection}. For ICM, topological features such as assortativity and radius appear to be selected for most values of $\gamma$ along with the in and out degree $\alpha$ estimation and various spectral eigenvalues. For uniform edge removal, average degree, clustering, diameter and radius are selected as well as betweenness centrality and some spectral eigenvalues. The MLE estimation of $\alpha$ is mostly absent, possibly due to high correlation with other features.

\section{Conclusion}
\label{sec_conclusion}

In the cat and mouse game of censorship and circumvention, sensitive word lists play a central role and are invaluable for measuring censorship \cite{fm2012}. However, in this paper we have shown that network structure is also a very promising avenue for measurement and detection of censorship. We examined the feasibility of automatically classifying networks as either censored or uncensored based on topological features. We compared two censorship strategies: (1) a uniform strategy where every post has an equal probability of being removed and (2) a strategy based on removing entire repost cascades. As expected, deletion of repost cascades was shown to result in higher classification accuracy. In the real world, however, models of censorship are far more complex and involve sensitive topics, users, as well as a combination of seemingly arbitrary post removals.

We identified salient topological properties including assortativity, average degree, deviations from scale-free degree distributions and average clustering coefficient that provide a starting point for exploring other local and global network features in the context of censorship detection.

There are some shortcomings in the present work. First, both the power law assumption and the configuration model for network generation are simplistic, so other degree distributions and network generators need to be examined. Second, we ignored the problem of sampling an online social network by directly generating the communication graphs. In reality, it not feasible to collect the complete communication graph due to the scale of the data. Thus, a future work will incorporate network sampling into the methodology to show how this affects classification. This is expected to negatively impact classifier accuracy. Third, the scale of the simulated networks size is small, with $|V|=1000$, however, we expect that for larger networks the features and subsequent classification results will stabilise, although at the cost of increased complexity. Finally, the methods presented in this preliminary study must be validated on real data. For this to be feasible it may be necessary to use different online social networks as sources of censored and uncensored reply-graphs.

There are several directions in which this work can be extended. Given that censorship primarily affects the diffusion of information, in addition to edge removal, we will examine how different levels of node censorship (i.e., suppression or removal of user accounts) affects the spread of information through an online social network. Second, the classification framework could be extended to provide a quantitative estimation of the level of censorship (i.e., the estimation of $\gamma$) in a given online social network.

\section{Acknowledgments}

This work was funded by the European Union (EU) and Science Foundation Ireland (SFI) in the course of the projects ROBUST (EU grant no. 257859) and CLIQUE Strategic Research Cluster (SFI grant no. 08/SRC/I1407), and LION-2 (SFI grant no. SFI/08/CE/I1380).

\bibliographystyle{acm}

\end{document}